\documentclass[11pt]{article}
\newcommand{\dis}{\displaystyle}

\title{{Lie point symmetries and first integrals: the Kowalevski top }}

\author{M. Marcelli and M. C. Nucci$^{a)}$}
\date{Dipartimento di Matematica
 e Informatica, Universit\`a di Perugia, 06123 Perugia, Italy.}
\begin{document}
 \maketitle
\renewcommand{\thefootnote}{$a)$}
\footnotetext{Corresponding author.
 E-mail: nucci@unipg.it}
\begin{abstract}
We show how the Lie group analysis method can be used in order to
obtain first integrals of any system of ordinary differential
equations.
 The method of reduction/increase of order
developed by Nucci ({J. Math. Phys.} {\bf 37}, 1772-1775 (1996))
is essential. Noether's theorem is neither necessary nor
considered. The most striking example we present is the
relationship between Lie group analysis and the famous first
integral of the Kowalevski top.
\end{abstract}
{\bf Running title:} Lie symmetries \& first integrals: Kowalevski
top
\newpage

\section{Introduction}
In January 2001, the first Whiteman prize for notable exposition
on the history of mathematics was awarded to Thomas Hawkins by the
American Mathematical Society. In the citation, published in the
Notices of AMS {\bf 48} 416-417 (2001), one reads that Thomas
Hawkins ``$\ldots$ has written extensively on the history of Lie
groups. In particular, he has traced their origins to [Lie's] work
in the 1870s on differential equations $\ldots$ the {\em id\'ee
fixe} guiding Lie's work was the development of a Galois theory of
differential equations $\ldots$ [Hawkins's book \cite{Hawkins2}]
highlights the fascinating interaction of geometry, analysis,
mathematical physics, algebra, and topology
$\ldots$".\\
In the Introduction of his book \cite{Olver},  Olver wrote that
``it is impossible to overestimate the importance of Lie's
contribution to modern science and mathematics. Nevertheless,
anyone who is already familiar with [it] $\dots$ is perhaps
surprised to know that its original inspirational source was the
field of differential equations".\\
Lie's monumental work on transformation groups  \cite{Lie 1},
\cite{Lie 2}, \cite{Lie 3}, and in particular contact
transformations \cite{Lie 4}, led him to achieve his goal
\cite{Lie 5}.
 Lie group
analysis is indeed the most powerful tool to find the general
solution of ordinary differential equations. Any known integration
technique can be shown to be a particular case of a general
integration method based on the derivation of the continuous group
of symmetries admitted by the differential equation, i.e. the Lie
symmetry algebra. In particular, Bianchi's theorem \cite{Bianchi},
\cite{Olver}, states that if an admitted $n$-dimensional solvable
Lie symmetry algebra is found, then the general solution of  the
corresponding  $n$ order system of ordinary differential equations
can be obtained by quadratures. The admitted Lie symmetry algebra
can be easily derived by a straightforward although lengthy
procedure. As  computer algebra softwares become widely used, the
integration of systems of ordinary differential equations by means
of  Lie group analysis is getting easier to carry out.   \\ A
major drawback of Lie's method is that it is useless when applied
to  systems of $n$ first order equations, because they admit an
infinite number of symmetries, and there is no systematic way to
find even one-dimensional Lie symmetry algebra, apart from trivial
groups like translations in time admitted by autonomous systems.
One may try to derive an admitted $n$-dimensional solvable Lie
symmetry algebra by making an ansatz on the form of its
generators. \\However, Nucci \cite{kepler} has remarked that  any
system of $n$ first order equations could be transformed into an
equivalent system where at least one of the equations is of second
order. Then, the admitted Lie symmetry algebra is no longer
infinite dimensional, and non trivial symmetries of the original
system could be retrieved \cite{kepler}. This idea has been
successfully applied in several instances \cite{kepler},
\cite{valenu}, \cite{core}, \cite{flu}, \cite{leitmann},
\cite{harmony}, \cite{classic},
\cite{lorpoin}.\\
Here we show another striking application of such an idea. If we
consider  a system of first  order equations, and by eliminating
one of the dependent variables derive an equivalent system which
has one equation of second order, then Lie group analysis applied
to that equivalent system yields the first integral(s)  of the
original system which does not contain the eliminated dependent
variable. Of course, if such first integrals exist.
  The procedure should be repeated on as many times as there are dependent
  variables  in order to find all such first integrals.
  The first integrals correspond to the characteristic curves of determining equations of
parabolic type \cite{smirnov} which are constructed by the method
of Lie group
analysis.\\
 We would like to remark that interactive (not automatic)
programs for calculating Lie point symmetries such as \cite{man1},
\cite{man2} are most appropriate for performing this task.
\\ It is well known that if one finds a
transformation which leaves invariant a functional describing a
variational problem, then Noether's theorem \cite{Noether}
provides a first integral of the corresponding Euler-Lagrange
system. Unfortunately,  a general method for finding such a
transformation does not exist. In addition, many equations of
physical interest (e.g., Lorenz system in meteorology
\cite{Lorenz}) do not come from a variational problem. On the
contrary, our method can be applied to any system of ordinary
differential equations, even if they do not derive from a
variational problem \cite{lorpoin}, and we do not make any a
priori hypothesis on the form of the first integrals, apart
missing one of the unknowns.
\\
In the next section, we describe the method in detail,  in
sections 3  and 4, we present the classical example of the
Kowalevski top, and in section 5 the three-dimensional Kepler
problem in cartesian coordinates. The last section contains some
final comments.
\section{Outline of the method}
Let us consider the following autonomous (which  could also be
non-autonomous) system of $N$ first order ordinary differential
equations
\begin{equation}
\left\{ \begin{array}{rcl}
        \dot w_1&=&F_1 (w_1,w_2,\ldots,w_N)\\
        \dot w_2&=&F_2 (w_1,w_2,\ldots,w_N)\\
        \vdots\\
        \dot w_N&=&F_N (w_1,w_2,\ldots,w_N)
\end{array} \right. \label{exm di sis} \end{equation}
Let
\begin{equation}
I=I(w_1, w_2, \ldots, w_{s-1}, w_{s+1}, \ldots, w_N),
\label{icon1}\\
\end{equation}
be a first integral  which does not depend on $w_s$, and
\begin{equation}
 X=V(t,w_1,\ldots,w_{N})\partial_t+\sum^{N}_{k=1}
G_k(t,w_1,\ldots,w_{N})\partial_{w_k} \label{simmetria dell'exm}
\end{equation}
be a generator of a Lie point symmetry group for (\ref{exm di
sis}). If we derive  $w_s$ from one of the equations (\ref{exm di
sis}), say the first, then we obtain a system of $N-2$ equations
of first order in $w_2,\ldots,w_{s-1},w_{s+1},\ldots,w_N$ and one
of second order in $w_1$. We remark that the method does not
depend on the equation we choose from (\ref{exm di sis}) to derive
$w_s$. After introducing the new notation $u_j, (j=1,\ldots,N-1)$,
we can write the system we obtain as
\begin{equation}
\left\{ \begin{array}{rcl}
        \ddot u_1&=&f_1 (u_1,u_2, \ldots,  u_{N-1}, \dot u_1)\\
        \dot u_2&=&f_2(u_1,u_2, \ldots,  u_{N-1}, \dot u_1) \\
        \vdots\\
                       \dot u_{N-1}&=&f_{N-1}(u_1,u_2, \ldots,  u_{N-1}, \dot u_1)
\end{array} \right. \label{exm di sis1} \end{equation}
A generator of a Lie point symmetry group for (\ref{exm di sis1})
is
\begin{equation}
 \bar X= \bar V(t,u_1,\ldots,u_{N-1})\partial_t+\sum^{N-1}_{j=1}
\bar G_j(t,u_1,\ldots,u_{N-1})\partial_{u_j} \label{gensis}
\end{equation}
If we apply Lie group analysis to system (\ref{exm di sis1}) using
the interactive REDUCE programs developed by Nucci \cite{man1},
\cite{man2}, then we obtain a determining equation of parabolic
type for $V$. Its characteristic curves will yield $m<N-1$
transformations,  which eliminate $\dot u_1$ from all the first
order equations in (\ref{exm di sis1}). Thus, we have obtained a
system of $N- 2$ equations of first order and one equation of
second order in the new dependent variables $\tilde u_j$ such that
$u_1=\tilde u_1$ and  each of the other variables $\tilde u_j$ are
 either the original $u_j$ itself, if $\dot u_1$ did not appear
in the $j$-equation of system (\ref{exm di sis1}), or the
corresponding characteristic curve. If we apply Lie group analysis
to this final system, then again a determining equation of
parabolic type will be derived, and its characteristic curve, when
rewritten in the original variables,
will be exactly the first integral (\ref{icon1}).\\
Now let us consider a system of $M$ second order ordinary
differential equations
\begin{equation}
\ddot x_i  = H_i (x_1,\ldots,x_M,\dot x_1,\ldots,\dot x_M),
\;\;\;\;(i=1,\ldots,M). \label{sysx}
\end{equation}
A generator of a Lie point symmetry group for this system has the
form
\begin{equation}
\Gamma=\tau(t,x_1,\ldots,x_{M})\partial_t+\sum^{M}_{i=1}
\eta_i(t,x_1,\ldots,x_{M})\partial_{x_i}. \label{gensyx}
\end{equation}
System (\ref{sysx}) can be converted into the following autonomous
system of $2M$ first order ordinary differential equations
\begin{equation}
\left\{ \begin{array}{rcl} \dot w_i & = &w_{M+i},\\
\dot w_{M+i} & = &H_i (w_1,\ldots,w_M,w_{M+1},\ldots,w_{2M}).
\end{array} \right. \label{syspw2}
\end{equation}
At this point, we could either proceed as indicated above or
choose one of the dependent variables to be the new independent
variable $y$ in order to reduce  the order of system
(\ref{syspw2}) by one \cite{kepler}. For example, we could take
$x_M\equiv w_M=y$. Then, system (\ref{syspw2}) becomes the
following non-autonomous system of $2M-1$ first order ordinary
differential equations  with independent variable $y$
\begin{equation}
\left\{ \begin{array}{rcl}
\displaystyle{{\rm d}\over{\rm d}y} w_h& = &{w_{M+h}/ w_{2M}}, \\
\displaystyle{{\rm d}\over {\rm d} y} w_{M+h}& = & {H_h
(w_1,\ldots,w_{M-1},y,w_{M+1},\ldots,w_{2M})/ w_{2M}},
 \end{array} \right. \label{syspwy2}
\end{equation}
where $(h=1,\ldots,M-1)$. Now, our method can be applied to this
system  as if it was system (\ref{exm di sis}). The fact that
system (\ref{syspwy2}) is not autonomous does not effect the
result, as we will show in the case of the three dimensional
Kepler problem in cartesian coordinates. \\
The same method can be applied to a single ordinary differential
equation of order $N$ which can be easily transformed into a
system of $N$ equations of first order. It should be noticed that
there could be several different ways of transforming an equation
of order $N$ into a system of $N$ equations of first order. Then,
the just described method may give different results, videlicet
(viz) no first integrals with certain reductions, all the first
integrals with different reductions.

\section{Finding the Kowalevski top}
The motion of a heavy rigid point about a fixed point is one of
the most famous problems of classical mechanics \cite{Golubev}. In
1750, Euler \cite{Euler} derived the equations of motion which now
bear his name, and described what is nowadays known as the
Euler-Poinsot case because of  the geometrical description given
by Poinsot about hundred years later \cite{Poinsot}.  It was
Jacobi \cite{Jacobi1} who integrated this case by using the
elliptic functions  which he had developed (along with Legendre,
Abel and Gauss \cite{Natucci}) and  mastered \cite{Jacobi2} (we
have translated this fundamental text into Italian and extensively
commented \cite{RNM}). Another case was described by Lagrange
\cite{Lagrange}, and it is known as the Lagrange-Poisson case, due
to the extensive study done later by Poisson \cite{Poisson} . This
case can also be integrated by using Jacobi elliptic functions
\cite{Whittaker}. At the time, it seemed that other cases could
easily be found and similarly integrated. In 1855, the Prussian
Academy of Science proposed this topic for a competition, but
nobody applied \cite{Cooke}. The problem was so elusive that the
German mathematicians called it the mathematical mermaid ({\em die
mathematische Nixe}) \cite{Kow91}. More than thirty years elapsed
before the Bordin prize was awarded to Kowalevski for finding and
reducing to hyperelliptic quadratures the third case
\cite{Kowalevski} which is since known as the Kowalevski top. She
solved the problem by looking for solutions which are
single-valued meromorphic functions in the entire complex plane of
the variable $t$ \cite{Golubev}. Her method became what is now
known as the Painlev\'e-Kowalevski
(or just Painlev\'e) method \cite{Ince}.\\
Hawkins  had established ``the nature and extent of Jacobi's
influence upon Lie" \cite{Hawkins1}. It is a remarkable
coincidence that
 the mathematical mermaid can also be found by using Lie group
analysis as we show in the following.\\
The Euler-Poisson equations describing the motion of a heavy rigid
body about a fixed point are \cite{Kowalevski}
\begin{equation}
 \left\{
    \begin{array}{rcl}
        \dot{p}&=&\Big((B-C)r
        q+mg(\beta z_{G}-\gamma y_{G})\Big)/A\\
        \dot{q}&=&\Big((C-A)p
        r+mg(\gamma x_{G}-\alpha z_{G})\Big)/B\\
        \dot {r}&=&\Big((A-B)p q+mg(\alpha y_{G}-\beta x_{G})\Big)/C\\
        \dot \alpha&=&\beta {r}-\gamma {q}\\
        \dot \beta&=&\gamma {p}-\alpha {r}\\
        \dot \gamma&=&\alpha {q}-\beta {p}
    \end{array}
 \right.\label{eulpois}
\end{equation}
with $A,B,C$ the principal moments of inertia , $p(t),q(t),r(t)$
the components of the angular velocity, $m$ the mass of the body,
$g$ the acceleration of gravity, $x_G,y_G,z_G$ the coordinates of
the center of mass, and $\alpha(t),\beta(t),\gamma(t)$ the
component of the unit vertical vector. There are three first
integrals for system (\ref{eulpois}):
 conservation of energy, i.e.
\begin{equation}
I_1=\frac{1}{2}\left(Ap^2+Bq^2+cr^2\right)+mg\,\left(x_G\alpha+y_G\beta+z_G\gamma
\right)\label{e0eu}\end{equation} conservation of the vertical
component of the angular momentum, i.e.
\begin{equation}I_2=Ap\alpha+Bq\beta+Cr\gamma\end{equation}
 the length of the unit vertical vector, i.e.
\begin{equation}I_3=\alpha ^2+\beta ^2+\gamma ^2  (=1)
\end{equation}
If we apply our method to system (\ref{eulpois}), then we find
only the first integral of the unit vertical vector which has
$p,q,r$ as missing variables.  Kowalevski found that  if one
imposes the following conditions on the parameters:
\begin{description}
\item{(1)} A=B=2C
\item{(2)} $z_{G}=0$, and either $x_{G}\neq 0$ or $y_{G}\neq 0$
 \end{description}
 then there exists a fourth
integral, i.e.
\begin{equation}
I_4=\left(p^2-q^2-mg\,\frac{x_G\alpha-y_G\beta}{C}\right)^2+
\left(2pq-mg\,\frac{x_G\beta+y_G\alpha}{C}\right)^2
\label{intekov}
\end{equation}
We notice that $\gamma$ and $r$ are missing in (\ref{intekov}).
Thanks to our method, we can find  the Kowalevski top by searching
for a first integral which does not contain $\gamma$. First we
derive $ \gamma $ from the second equation of system
(\ref{eulpois}),
 i.e.
$$ \gamma =\dis\frac{B\dot {q}+(A-C)pr+mgz_G\alpha}{mgx_{G}}$$
which implies that $x_G$ must be different from zero. We obtain
the following  system
 of four equations of first order, and one of second order:
\begin{equation}
           \begin{array}{rcl}

\ddot u_{1}&=&\dot u_1\left(Au_1 z_G +
(A-C)u_3y_G\right)/Ax_G\\&&- (A - B)(A - C)u_1u_2^2/BC +(A -
C)^2y_Gu_2u_3^2/ABx_G\\&&  +(A - C)u_1u_2u_3z_G/Bx_G- (A - C)(B
-C)u_1u_3^2/AB
\\&& - (Au_2x_G - Cu_3z_G)(A - C)mgy_Gu_4/ABCx_G\\&&
+\left(A(A-2C)u_2x_G+C(C-2A)u_3z_G\right)mgu_5/ABC\\&&+(x_G^2 +
z_G^2)mgu_1u_4/Bx_G
\end{array}\label{sis1eu1}
\end{equation}
\begin{equation}
          \begin{array}{rcl}

   \dot u_{2}&=&- \dot u_1By_G/Ax_G+u_3\left(( C- A)y_Gu_2+(B - C)x_Gu_1\right)/Ax_G\\
   && +mgz_G( - u_4y_G + u_5x_G)/Ax_G
   \end{array}\label{sis1eu2}
\end{equation}
\begin{equation}
\begin{array}{rcl}

          \dot u_{3}&=&\left((A-B)u_1u_2 + mg(u_4y_G - u_5x_G)\right)/C
          \end{array}\label{sis1eu3}
\end{equation}
\begin{equation}
\begin{array}{rcl}

          \dot u_{4}&=&- \dot u_1Bu_1/mgx_G+(C
          -A)u_1u_2u_3/mgx_G\\
                    &&+( u_3u_5x_G- u_1u_4z_G )/x_G
                    \end{array}\label{sis1eu4}
\end{equation}
\begin{equation}
\begin{array}{rcl}

          \dot u_{5}&=&\dot u_1Bu_2/mgx_G+(A -
          C)u_2^2u_3/mgx_G+z_Gu_2u_4/x_G-u_3u_4
      \end{array}
    \label{sis1eu5}
\end{equation} with \begin{equation} u_{1}={q},\; u_{2}={p},\;
u_{3}={r},\; u_{4}=\alpha ,\; u_{5}=\beta \label{psosteu}
\end{equation}
Now we apply Lie group analysis to system
(\ref{sis1eu1})-(\ref{sis1eu5}). An operator $\Gamma$
\begin{equation}\Gamma=V
(t,u_{1},u_{2},u_{3},u_4,u_5)\partial_t+\sum_{k=1}^5G_{k}
(t,u_{1},u_{2},u_{3},u_4,u_5)\partial_{u_{k}}\end{equation} is
said to generate a Lie point symmetry group if its second
prolongation
$${\displaystyle{\Gamma} \atop {\rm {\small{2}}}}=\Gamma+\sum_{k=1}^5\left({{\rm d}G_k\over{\rm d}t}
-\dot u_k{{\rm d}V\over{\rm d}t}\right)\partial_{\dot
u_k}+\left({{\rm d}\over {\rm d}t}\left({{\rm d}G_1\over{\rm d}t}
-\dot u_1{{\rm d}V\over{\rm d}t}\right)-\ddot u_1{{\rm
d}V\over{\rm d}t}\right)\partial_{\ddot u_1}$$ applied to system
(\ref{sis1eu1})-(\ref{sis1eu5}), on their solutions, is
identically equal to zero, i.e.
\begin{equation}
\begin{array}{rcl}

\displaystyle{\displaystyle{\Gamma} \atop
 \small{2}}(\ref{sis1eu1}){\Big
|}_{(\ref{sis1eu1})-(\ref{sis1eu5})}&=&0 \\
\displaystyle{\displaystyle{\Gamma} \atop
 \small{2}}(\ref{sis1eu2}){\Big
|}_{(\ref{sis1eu1})-(\ref{sis1eu5})}&=&0 \\
\displaystyle{\displaystyle{\Gamma} \atop
 \small{2}}(\ref{sis1eu3}){\Big
|}_{(\ref{sis1eu1})-(\ref{sis1eu5})}&=&0 \\
\displaystyle{\displaystyle{\Gamma} \atop
 \small{2}}(\ref{sis1eu4}){\Big
|}_{(\ref{sis1eu1})-(\ref{sis1eu5})}&=&0 \\
\displaystyle{\displaystyle{\Gamma} \atop
 \small{2}}(\ref{sis1eu5}){\Big
|}_{(\ref{sis1eu1})-(\ref{sis1eu5})}&=&0 \\
\end{array}\label{deteq5}
\end{equation}
The five determining equations (\ref{deteq5}) constitute an
overdetermined system of linear partial differential equations in
the unknowns $V, G_{k} (k=1,5)$ In fact, they are polynomials in
$\dot u_1$, each coefficient of which must become identically
equal to zero.
 In particular, the first determining equation in (\ref{deteq5}) is a polynomial of degree three for
$\dot u_1$.  The coefficient of highest degree  yields an equation
of parabolic type for $V$ in four independent variables
$u_1,u_2,u_4,u_5$, i.e.
\begin{eqnarray}
A^2mg^2x_G^2\displaystyle\frac{\partial^2 V}{\partial {u_{1}^2}}-
2ABm^2g^2x_Gy_G\displaystyle\frac{\partial^2 V}{\partial {u_{1}
u_{2}}}- 2A^2Bmgu_1x_G\displaystyle\frac{\partial^2 V}{\partial
{u_{1}
u_{4}}}\nonumber\\+2A^2Bmgu_2x_G\displaystyle\frac{\partial^2
V}{\partial {u_{1}
u_{5}}}+B^2m^2g^2y_G^2\displaystyle\frac{\partial^2 V}{\partial
{u_{2}^2}}+2AB^2mgu_1y_G\displaystyle\frac{\partial^2 V}{\partial
{u_{2} u_{4}}}\\- 2AB^2mgu_2y_G\displaystyle\frac{\partial^2
V}{\partial {u_{2}
u_{5}}}+A^2B^2u_1^2\displaystyle\frac{\partial^2 V}{\partial
{u_{4}^2}}- 2A^2B^2u_1u_2\displaystyle\frac{\partial^2 V}{\partial
{u_{4} u_{5}}}\nonumber
\\+A^2B^2u_2^2\displaystyle\frac{\partial^2 V}{\partial
{u_{5}^2}}-A^2Bmgx_G\displaystyle\frac{\partial V}{\partial
{u_{4}}}- AB^2mgy_G\displaystyle\frac{\partial V}{\partial
{u_{5}}}=0\nonumber
\end{eqnarray}

Its three characteristic curves yield the following
transformations
\begin{equation}
\begin{array}{rcl}

u_{2}=s_2-\dis\frac{Bu_1y_G}{Ax_G} &{\rm viz} &
p=s_2-\dis\frac{Bqy_G}{Ax_G}  \\
 u_{4}=s_4-\dis\frac{Bu_1^2}{2mg x_G} &{\rm viz} &
\alpha
=s_4-\dis\frac{Bq^2}{2mg x_G}  \\
u_{5}=s_5+Bu_1\dis\frac{By_Gu_1+2Ax_Gu_2}{2Amgx_G^2} &{\rm viz} &
\beta =s_5+Bq\dis\frac{By_Gq+2Ax_Gp}{2Amgx_G^2}
\end{array} \label{firsttreu} \end{equation}
with $s_2$, $ s_{4} $, and $s_{5} $  new unknown functions of $ t
$.\\
 As outlined in section 2, transformations (\ref{firsttreu})
 eliminate $\dot u_1$ from all the first order
equations in system (\ref{sis1eu1})-(\ref{sis1eu5}).\\ In fact,
system (\ref {sis1eu1})-(\ref {sis1eu5}) becomes:
\begin{equation}
       \begin{array}{rcl}

 \ddot u_1 &=&
 ( - 6 A^{2} B C   {u_1}  {\tilde{u}_2}^{2} {x_G}^{2}
- 4 A^{2} B C   {u_1}  {x_G}  {\tilde{u}_2}  {u_3}  {z_G}
 \\&&- 2 A^{2} B  {mg}  {u_1}  {x_G}^{2}
 {y_G}  {\tilde{u}_5}
 - 2 A^{2} C^{2}  {u_3}  {x_G}  {u_1}   {\tilde{u}_2}  {z_G}\\&&  + 2 A^{2} B
{mg}  {u_1 }  {x_G}  {y_G}^{2}  {\tilde{u}_4}- 3 A^{2} B C
 {u_1}^{2}  {y_G}  {u_3}  {z_G}\\&&
 + 2 A^{2} B C  {\dot u_1}  {u_3}  {x_G}   {y_G} - 5 A^{2} B C
{u_1}^{2}  {y_G}   {\tilde{u}_2}  {x_G} + 3 A B ^{3}  {u_1}^{3}
{y_G}^{2}\\&& - 2 A^{2} B C  {u_1} { u_3}^{2} {y_G}^{2}  - A^{2} B
C  {u_1}^{3}  {x_G}^{2} - 2 A^{2 } B C {u_1} {u_3}^{2}  {x_G}^{2}
\\&&- 2 A C^{3} {u_3} ^{2}  {x_G}^{2}  {u_1}   + 2 A^{2} C  {mg}
{x_G} {u_1}
 {\tilde{u}_4}  {z_G}^{2} + 2 A^{2} C  {mg}  {
x_G}^{3}  {u_1}  {\tilde{u}_4} \\&&- 4 A^{2} C  {mg}
 {x_G}^{2}  {u_3}  {z_G}  {\tilde{u}_5}
 - 4 A^{2} C  {mg}  {x_G}^{3}  {\tilde{u}_2}   {\tilde{u}_5} \\&&+ 2 A^{2} C {mg}
{x_G}  {u_3 }  {z_G}  {\tilde{u}_4}  {y_G} - 7 A^{2} B^{2}
 {y_G}  {u_1}^{2}  {\tilde{u}_2}  {x_G}
\\&& + 2 A^{2} C  {mg}  {x_G}^{2}  {\tilde{u}_2}   {\tilde{u}_4}  {y_G} - 4 A^{2}
C^{2}  {u_3}^{2}   {x_G}  {\tilde{u}_2}  {y_G} \\&&  - 3 A^{2}
B^{2}  { y_G}^{2} {u_1}^{3} - B^{2} C^{2}  {y_G}  {u_1}^{2}  {u_3}
{z_G} - 2 {u_1} A^{4} {\tilde{u}_2}^{2}  { x_G}^{2} \\&&+ 2 A^{3}
{\tilde{u}_2}  {x_G} C {u_3}^{2}
  {y_G}
 + 2 A^{3}  {\tilde{u}_2}  {x_G} C  {u_1}
 {u_3}  {z_G} + 2 A^{3}  {\tilde{u}_2}^{2}  {x_G}
^{2} C  {u_1} \\&&- 2  {u_1} B C^{3}  {y_G}^{2}
  {u_3}^{2}
 + 5 A^{3}  {\tilde{u}_2}  {x_G} B  {u_1}^{2}   {y_G} - 2 A^{3} {\tilde{u}_2}
{x_G}^{2}  { mg}  {\tilde{u}_4}  {y_G} \\&&+ 4 A^{3}
{\tilde{u}_2}^{2}
 {x_G}^{2} B  {u_1}
 + 4 A B C  {mg}  {u_1}  {x_G}^{2}
 {y_G}  {\tilde{u}_5}\\&& - 2 A B C  {mg}  {u_1}
 {x_G}  {y_G}^{2}  {\tilde{u}_4} - 2 A B C^{2}
 {\dot u_1}  {u_3}  {x_G}  {y_G}
 \\&&+ 2 A B C^{2}  {u_1}  {x_G}  {\tilde{u}_2}
 {u_3}  {z_G} + 3 A B C^{2}  {u_1}^{2}
 {y_G}  {u_3}  {z_G} \\&&  + 4 A B C^{2}  {
u_1}  {u_3}^{2}  {y_G}^{2} + 2 A B C^{2}  {u_1}  {u_3}^{2} {
x_G}^{2} + 2 A B^{2} C {u_1}^{2}  {y_G}  { u_3}  {z_G} \\&&+ 10 A
B^{2} C {u_1}^{2}  {y_G}
  {\tilde{u}_2}  {x_G}
 + 2 A C^{3}  {u_3}^{2}  {x_G}  {\tilde{u}_2}   {y_G} \\&&+ 2 A C^{2} {mg}
{u_3}  {x_G }^{2}  {z_G}  {\tilde{u}_5} - 4 B^{3} C  {y_G}^{2}
 {u_1}^{3}
 \\&&- 2 A C^{2}  {mg}  {u_3}  {x_G}
 {z_G}  {\tilde{u}_4}  {y_G} + 2 A^{2} C^{2}  {
u_3}^{2}  {x_G}^{2}  {u_1} \\&&- A^{2} B C  {u_1} ^{3}  {z_G}^{2}
+ 3 A B^{2} C  {u_1}^{3}  { y_G}^{2} \\&&  + 2 A^{2} B C  {u_1}
{x_G} {\dot u_1} {z_G} + 2 A^{3}  {\tilde{u}_2} {x_G}^{3}  { mg}
{\tilde{u}_5})/2A^{2} B C {x_G}^{2}
\end{array}\label{sis2eu1}
\end{equation}

\begin{equation}
          \begin{array}{rcl}

 \dot {\tilde{u}}_{2} &=&
( - 2 A^{2}   {\tilde{u}_2}  {u_3} {x_G} {y_G} + A B
 {u_1}^{2}  {y_G}  {z_G} + 2 A B  {u_1}
  {\tilde{u}_2}  {x_G}  {z_G}\\&& + 2 A B  {u_1}
 {u_3}  {x_G}^{2}
 + 2 A B  {u_1}  {u_3}  {y_G}^{2} - 2  A C  {u_1}  {u_3}
{x_G}^{2} \\&&+ 2 A C
 {\tilde{u}_2}  {u_3}  {x_G}  {y_G} - 2 A
 {mg}  {\tilde{u}_4}  {x_G}  {y_G}  {z_G}
 + 2 A  {mg}  {\tilde{u}_5}  {x_G}^{2}   {z_G} \\&&- 2 B C  {u_1}  {u_3}
{y_G}^{2}) /2A^{2}  {x_G}^{2}

\end{array}\label{sis2eu2}
\end{equation}
\begin{equation}
          \begin{array}{rcl}

 {\dot u_3} &=&  (2 A^{2}  {
u_1}  {\tilde{u}_2}  {x_G} - 3 A B  {u_1}^{2}   {y_G} - 4 B  {u_1}
{\tilde{u}_2} A {x_G} + 2  A  {mg}  {\tilde{u}_4}  {x_G} {y_G}\\&&
- 2
  {\tilde{u}_5}  {x_G}^{2} A  {mg}
 + 3 B^{2}  {y_G}  {u_1}^{2})/2A C  { x_G}
\end{array}\label{sis2eu3}
\end{equation}
\begin{equation}
          \begin{array}{rcl}

 \dot {\tilde{u}}_{4} &=&  ( - 2 A^{2}
 {u_1}  {\tilde{u}_2}  {u_3}  {x_G} + A B
 {u_1}^{3}  {z_G} + 2 A B  {u_1}^{2}  {
u_3}  {y_G} \\&&+ 2 A B  {u_1}  {\tilde{u}_2}  {u_3}
  {x_G}
 + 2 A C  {u_1}  {\tilde{u}_2}  {u_3}   {x_G} - 2 A  {mg}  {u_1} {\tilde{u}_4}
{x_G}  {z_G} \\&&+ 2 A  {mg}  {u_3}
 {\tilde{u}_5}  {x_G}^{2} - B^{2}  {u_1}^{2}  {u_3
}  {y_G}
 - 2 B C  {u_1}^{2}  {u_3}  {y_G})/2A   {mg}  {x_G}^{2}
 \end{array}\label{sis2eu4}
\end{equation}
\begin{equation}
          \begin{array}{rcl}

 \dot {\tilde{u}}_{5} &=&  (2 A^{2}  {
\tilde{u}_2}^{2}  {u_3}  {x_G} - A B  {u_1}^{2}   {\tilde{u}_2}
{z_G} + A B {u_1}^{2} {u_3}   {x_G} - 2 A B  {u_1}  {\tilde{u}_2}
{u_3} {y_G} \\&&
 - 2 A C  {\tilde{u}_2}^{2}  {u_3}  {x_G} + 2  A  {mg}  {\tilde{u}_2}  {\tilde{u}_4}
{x_G}  { z_G} - 2 A  {mg}  {u_3}  {\tilde{u}_4}  {x_G}^{2 } \\&&-
2 B^{2} {u_1}^{2} {\tilde{u}_2}  {z_G}   - 2 B^{2}  {u_1}^{2}
{u_3} {x_G} + 2 B C {u_1}^{2} {u_3} {x_G} \\&&+ 2 B C
 {u_1}  {\tilde{u}_2}  {u_3}  {y_G} - 2 B
 {mg}  {u_1}  {\tilde{u}_5}  {x_G}  {z_G})/
2A {mg}  {x_G}^{2}
 \end{array}       \label{sis2eu5}
\end{equation}

 with \begin{equation}
\tilde{u}_{2}=s_{2}, \;\;\;\tilde{u}_{4}=s_{4},
\;\;\;\tilde{u}_{5}=s_{5} \label{secsosteu}
\end{equation}
 We now apply Lie group analysis
to system (\ref{sis2eu1})-(\ref{sis2eu5}). An operator
$\tilde{\Gamma}$
\begin{equation}
\begin{array}{rcl}

\tilde\Gamma&=&\tilde V (t,u_{1},\tilde u_{2},u_{3},\tilde
u_4,\tilde u_5)\partial_t+\tilde G_{1} (t,u_{1},\tilde
u_{2},u_{3},\tilde u_4,\tilde u_5)\partial_{u_{1}}\\&&+\tilde
G_{2} (t,u_{1},\tilde u_{2},u_{3},\tilde u_4,\tilde
u_5)\partial_{\tilde u_{2}}+\tilde G_{3} (t,u_{1},\tilde
u_{2},u_{3},\tilde u_4,\tilde u_5)\partial_{u_{3}}\\
&&+\tilde G_{4} (t,u_{1},\tilde u_{2},u_{3},\tilde u_4,\tilde
u_5)\partial_{\tilde u_{4}}+\tilde G_{5} (t,u_{1},\tilde
u_{2},u_{3},\tilde u_4,\tilde u_5)\partial_{\tilde
u_{5}}\end{array}\end{equation}  is said to generate a Lie point
symmetry group if its second prolongation
$\displaystyle{\tilde{\displaystyle{\Gamma} \atop  \small{2}}}$
 applied to system
(\ref{sis2eu1})-(\ref{sis2eu5}), on their solutions, is
identically equal to zero, i.e.
\begin{equation}
\begin{array}{rcl}

\displaystyle{\tilde{\displaystyle{\Gamma} \atop
 \small{2}}}(\ref{sis2eu1}){\Big
|}_{(\ref{sis2eu1})-(\ref{sis2eu5})}&=&0 \\
\displaystyle{\tilde{\displaystyle{\Gamma} \atop
 \small{2}}}(\ref{sis2eu2}){\Big
|}_{(\ref{sis2eu1})-(\ref{sis2eu5})}&=&0 \\
\displaystyle{\tilde{\displaystyle{\Gamma} \atop
 \small{2}}}(\ref{sis2eu3}){\Big
|}_{(\ref{sis2eu1})-(\ref{sis2eu5})}&=&0 \\
\displaystyle{\tilde{\displaystyle{\Gamma} \atop
 \small{2}}}(\ref{sis2eu4}){\Big
|}_{(\ref{sis2eu1})-(\ref{sis2eu5})}&=&0 \\
\displaystyle{\tilde{\displaystyle{\Gamma} \atop
 \small{2}}}(\ref{sis2eu5}){\Big
|}_{(\ref{sis2eu1})-(\ref{sis2eu5})}&=&0 \\
\end{array}\label{det2eq5}
\end{equation}
The five determining equations (\ref{det2eq5}) constitute an
overdetermined system of linear partial differential equations in
the unknowns $\tilde V, \tilde G_{k} (k=1,5)$ In fact, they are
polynomials in $\dot u_1$, each coefficient of which must become
identically equal to zero. In particular, the fifth determining
equation in (\ref{det2eq5}) is a polynomial of degree one for
$\dot u_1$. We call its two coefficients $c5k1$ and $c5k0$. For
the sake of simplicity, we assume $\tilde
G_k=0,\displaystyle{\partial \tilde V \over \partial t}=0$. Then,
the coefficient of degree one, i.e. $c5k1$,  yields
$${\partial \tilde V \over \partial u_1}=0$$
Now, $c5k0$ is a polynomial of degree five in $u_1$. Therefore,
its coefficients, call them $c5m5, c5m4, c5m3,c5m2,c5m1,c5m0$,
must become identically equal to zero. The coefficient of degree
five in $u_1$, i.e. $c5m5$, yields
\begin{equation}
{\partial \tilde V\over \partial \tilde u_4}A^2B^2Cz_G\left( -
(A+2B)\tilde u_2z_G + (A-2B+2C)u_3x_G \right)=0
\end{equation}
which gives the condition on the parameter \begin{equation}
z_G=0\label{1cond}\end{equation} Then, the coefficient of degree
four in $u_1$, i.e. $c5m4$, yields
\begin{eqnarray}
 - \left(3(A - B){\partial \tilde V\over \partial u_3}mg y_G x_G - (A - 2B +2C)
{\partial \tilde V\over \partial \tilde u_5} Cu_3x_G \right .\nonumber\\
\left .- (2A - B -2C){\partial \tilde V\over \partial \tilde
u_4}Cu_3y_G\right)(A - 2B + 2C)AB^2u_3x_G=0\label{c5m4}
\end{eqnarray}
which gives the condition on the parameters
\begin{equation}A=2B-2C\label{2cond}\end{equation}
Then, the coefficient of degree three in $u_1$, i.e. $c5m3$,
becomes
\begin{equation}
12\left({\partial \tilde V\over \partial u_3}mgx_G - {\partial
\tilde V\over \partial \tilde u_4}Cu_3\right)(2B - 3C)(B - C)(B
-2C)B^2\tilde u_2u_3y_G^2=0\end{equation} which gives the further
condition on the parameters
\begin{equation}B=2C\label{3cond}\end{equation}
Thus, we have found the Kowalevski top.  We also notice that
either condition $2B=3C$ or $B=C$ leads to the Lagrange top.
Finally, we are left with two linear first order partial
differential equations in $\tilde V=\tilde V(\tilde u_2,u_3,\tilde
u_4,\tilde u_5)$, the coefficient of degree  two in $u_1$, i.e.
$c5m2$,
\begin{equation}
2 \left({\partial \tilde V\over \partial \tilde u_4}x_G -
{\partial \tilde V\over \partial \tilde u_5}y_G\right)Cu_3 \tilde
u_2- 4{\partial \tilde V\over \partial u_3}mgx_G^2\tilde u_2 + (
x_G^2 + y_G^2){\partial \tilde V\over \partial \tilde
u_2}mgu_3=0\label{c5m2}\end{equation} and the coefficient of
degree  one in $u_1$, i.e. $c5m1$
\begin{eqnarray}
2C\left(C\tilde u_2^2 - mg\tilde u_4x_G - mg\tilde
u_5y_G\right){\partial \tilde V\over \partial \tilde
u_4}x_Gu_3\tilde u_2\nonumber\\ -4C\left(C\tilde u_2^2-mg\tilde
u_4x_G\right){\partial \tilde V\over\partial \tilde
u_5}y_Gu_3\tilde u_2 \label{c5m1}\\- 2\left(2C\tilde u_2^2x_G -
2mg\tilde u_4x_G^2 + mg\tilde u_4y_G^2 - mg\tilde
u_5x_Gy_G\right){\partial \tilde V\over \partial u_3}mgx_G\tilde
u_2 \nonumber\\+ \left(C\tilde u_2^2x_G^2 + 2C\tilde u_2^2y_G^2 -
mg\tilde u_4x_G^3 - mg\tilde u_4x_Gy_G^2\right){\partial \tilde
V\over
\partial \tilde u_2}mgu_3=0 \nonumber
\end{eqnarray}
If $\tilde V$ satisfies equations (\ref{c5m2}) and (\ref{c5m1}),
then it is easy to prove that the determining equations
(\ref{det2eq5}) are identically satisfied by considering
conditions (\ref{1cond}), (\ref{2cond}), (\ref{3cond}) as well.\\
From (\ref{c5m2}) it is easy to obtain that $ \tilde V = \tilde
V(\eta_1, \eta_2, \eta_3) $ with
\begin{equation}
\eta_{1}=u_3^2+\displaystyle{4x_G^2 \tilde u_2^2\over
x_G^2+y_G^2},\;\; \eta_2= \tilde u_4-\displaystyle{Cx_G\tilde
u_2^2\over mg(x_G^2+y_G^2)},\;\; \eta_3 =\tilde
u_5+\displaystyle{Cy_G\tilde u_2^2\over mg(x_G^2+y_G^2)}
\label{etas}
  \end{equation}Then,
(\ref{c5m1}) becomes \begin{equation}
 2mg(y_G\eta_2-x_G\eta_3){\partial \tilde V\over\partial
\eta_1}+C{\partial \tilde V\over\partial \eta_2}\eta_3-C{\partial
\tilde V\over\partial \eta_3}\eta_2 =0 \label{c5m1.2}
\end{equation}
 Its characteristic curves are
\begin{equation}
\xi_1=\eta_{1}+\dis{2mg\over C}(y_G\eta_3+x_G\eta_2),
\;\;\;\;\xi_2=\eta_{2}^2+\eta_{3}^2 \label{xiskov}
\end{equation} Finally, we have that $ \tilde V=\Psi(\xi_1,\xi_2)$ with $\Psi$ an arbitrary
function of $\xi_1,\xi_2$, and consequently operator
\begin{equation}
\tilde \Gamma=\Psi(\xi_1,\xi_2){\partial}_t
\label{opkov}\end{equation} is a generator of a Lie point symmetry
for system (\ref{sis2eu1})-(\ref{sis2eu5}). Transforming
(\ref{xiskov}) into the original unknown functions by using
(\ref{etas}), (\ref{secsosteu}), (\ref{firsttreu}),
(\ref{psosteu}) yields
$$\xi_{1}=\dis\frac{2}{C}\left(\dis\frac{C}{2}
\left( 2{p}^2+2{q}^2+{r}^2 \right)+mg\,(x_G\alpha+y_G\beta)\right)
$$
$$
\xi_{2}=C^2\dis{\left(p^2-q^2-mg\,\dis\frac{x_G\alpha-y_G\beta}{C}\right)^2+
\left(2pq-mg\,\dis\frac{x_G\beta+y_G\alpha}{C}\right)^2\over
m^2g^2(x_G^2+y_G^2)}$$ which correspond to
 the first integral of conservation of energy (\ref{e0eu}),
and
 that derived by Kowalevski (\ref{intekov}), respectively.\\
Can other cases of integrability (viz integration by quadrature)
be obtained by using our method?  We leave the answer to a future
paper. In \cite{valenu}, the application of our method led to an
integrable case for a nonlinear system of three ordinary
differential equations which does not possess the Painlev\'e
property.
\section{First integrals of the Kowalevski top}
We apply our method to the
Kowalevski top itself which corresponds to the  following
conditions on the parameters:
\begin{description}
\item{(1)} A=B=2C
\item{(2)} $y_{G}=z_{G}=0, \qquad  x_{G}>0$
 \end{description}
 The condition on $y_G$ can be added without loss of generality.
 Then, system (\ref{eulpois}) become
\begin{equation}
 \left\{
    \begin{array}{rcl}
        \dot {p}&=&{r q}/{2}\\
        \dot{q}&=&-{p r}/2+mgx_{G}\gamma/{2C} \\
        \dot {r}&=&-{mg x_{G}}\beta /{C}\\
        \dot \alpha&=&\beta {r}-\gamma {q}\\
        \dot \beta&=&\gamma {p}-\alpha {r}\\
        \dot \gamma&=&\alpha {q}-\beta {p}\\
    \end{array}
 \right.
 \label{sisoriginario}
\end{equation}
The first integrals for the Kowalewski top are
\begin{description}
\item{(1)} conservation of energy, i.e.
\begin{equation} I_{1}=\dis\frac{C}{2} \left(2
{p}^2+2{q}^2+{r}^2 \right)+mg\,x_{G}\alpha \label{eo}
\end{equation}
\item{(2)} conservation of the vertical component of the angular
momentum, i.e.
\begin{equation} I_{2}=C(2{p}\alpha +2{q}\beta
+{r}\gamma ) \label{ivert}
\end{equation}
\item{(3)} the length of the unit vertical vector, i.e.
 \begin{equation} I_{3}=\alpha ^2+\beta ^2+\gamma ^2  (=1)
\label{igenerale}
\end{equation} \item{(4)} the first integral derived by Kowalevski, i.e.
 \begin{equation} I_{4}=\left({p}^2-{q}^2-\dis\frac{x_{G}\alpha
mg}{C}\right)^2+ \left(2{p q}-\dis\frac{x_{G}\beta mg}{C}\right)^2
\label{ikov}
\end{equation}
\end{description}
If our method is applied to (\ref{sisoriginario}), then all the
first integrals can be obtained, apart from (\ref{ivert}) which
has all the unknown variables $p,q,r,\alpha,\beta,\gamma$
appearing in its expression. Let us observe that
\begin{itemize}
\item
 $\beta$ does not appear in $I_1$
\item $\gamma$ does not appear in both $I_1$ and $I_4$
\item $p$ does not appear in $I_3$
\end{itemize}
In the following, we eliminate  $\alpha,\beta,\gamma, p$ from
system (\ref{sisoriginario}) one at a time.
\subsection{Eliminating $\alpha$} First we show a negative result: no first
integral obtained. Let us assume that we do not know any of the
first integrals. Therefore, we do not know a priori that none of
the first integrals can be obtained by deriving $\alpha$.
 We derive  $ \alpha $ from the fifth equation of system (\ref{sisoriginario}),
 i.e.
$$\alpha =\dis\frac{{p}\gamma-\dot \beta}{{r}}$$ and obtain the following  system
 of four equations of first order, and one of second order:
 \begin{equation}
    \left\{
       \begin{array}{ll}
          \ddot u_{1}=(- 2Cu_{1}u_{3}^3 - 2Cu_{1}u_{3}u_{4}^2 - 2C\dot u_{1}u_{2}u_{4} +
          3Cu_{2}u_{3}^2u_{5} \\ \qquad + 2Cu_{2}u_{4}^2u_{5}-2mgu_{1}\dot u_{1} x_{G}  +
          2mgu_{1}u_{4}u_{5} x_{G} )/2Cu_{3}\\
          \dot u_{2}=(- Cu_{3}u_{4} + mgu_{5} x_{G})/{2C}\\
          \dot u_{3}=(- mgu_{1} x_{G} )/{C}\\
          \dot u_{4}=(u_{2}u_{3})/{2}\\
          \dot u_{5}=(- u_{1}u_{3}u_{4} - \dot u_{1}u_{2} + u_{2}u_{4}u_{5})/{u_{3}}
      \end{array}
    \right.
    \label{sis1}
\end{equation} with \begin{equation} u_{1}=\beta,\; u_{2}={q},\;
u_{3}={r},\; u_{4}={p},\; u_{5}=\gamma \label{prima sost }
\end{equation}
If we apply Lie group analysis to system (\ref{sis1}), then we
obtain a determining equation of parabolic type for $V$ in two
independent variables. Its characteristic curve is
$$u_5 u_3+u_1u_2$$
 which yields the following transformation
\begin{equation} u_{5}=\dis\frac{s_{5}-u_{2}u_{1}}{u_{3}}\;\;\;\; {\rm
viz}\;\;\;\; \gamma =\dis\frac{s_{5}-{q}\beta}{{r}}\label{cc}
\end{equation} with $s_{5}$ a new unknown function of
$ t $. Then, system   (\ref {sis1}) transforms into:
 \begin{equation}
    \left\{
       \begin{array}{ll}
          \ddot u_{1}=(- 3Cu_{1}u_{2}^2u_{3}^2 - 2Cu_{1}u_{2}^2u_{4}^2
 - 2Cu_{1}u_{3}^4- 2Cu_{1}u_{3}^2u_{4}^2  \\ \qquad
\;\;\;  - 2C\dot u_{1}u_{2}u_{3}u_{4}  +
3Cu_{2}u_{3}^2\tilde{u}_{5} + 2Cu_{2}u_{4}^2\tilde{u}_{5} -
          2mgu_{1}^2u_{2}u_{4} x_{G}  \\ \qquad \;\;\;- 2mgu_{1}\dot u_{1}u_{3} x_{G}  +
          2mgu_{1}u_{4}\tilde{u}_{5} x_{G})/{2Cu_{3}^2}\\
          \dot u_{2}=(- Cu_{3}^2u_{4} - mgu_{1}u_{2} x_{G}  + mg\tilde{u}_{5} x_{G} )/{2Cu_3}\\
          \dot u_{3}=(- mgu_{1} x_{G} )/{C}\\
          \dot u_{4}=u_{2}u_{3}/{2}\\
          \dot {\tilde{u}}_{5}=( - 2Cu_{1}u_{2}^2u_{4} - 3Cu_{1}u_{3}^2u_{4} +
          2Cu_{2}u_{4}\tilde{u}_{5}+ mgu_{1}^2u_{2} x_{G}\\ \qquad \;\;\;  -
          mgu_{1}\tilde{u}_{5} x_{G})/{2Cu_{3}}
      \end{array}
    \right.
    \label{sis2}
\end{equation} with \begin{equation}
\tilde{u}_{5}=s_{5} \label{seconda sost}
\end{equation} If we apply Lie group analysis to system (\ref{sis2}), then we obtain a
two dimensional Lie symmetry algebra generated by the following
two operators:
\begin{eqnarray} \Gamma_{1}&=&\dis\frac{\partial}{\partial t},
\label{det} \\ \Gamma_{2}&=&-t\dis \frac{\partial}{\partial t}+
2u_{1}\dis \frac{\partial}{\partial u_{1}}+ u_{2}\dis
\frac{\partial}{\partial u_{2}}+u_{3}\dis \frac{\partial}{\partial
u_{3}}+u_{4}\dis \frac{\partial}{\partial u_{4}} \nonumber \\ &&
+3\tilde{u}_{5}\dis \frac{\partial}{\partial \tilde{u}_{5}}
\label{scaling}
\end{eqnarray}  which in the original unknown functions correspond to:
\begin{eqnarray} \Gamma_{1}&=&\dis\frac{\partial}{\partial t}
\label{det1} \\ \Gamma_{2}&=&-t\dis \frac{\partial}{\partial t}+
p\dis \frac{\partial}{\partial p}+ q\dis \frac{\partial}{\partial
q}+r\dis \frac{\partial}{\partial r}+2\alpha\dis
\frac{\partial}{\partial \alpha} \nonumber \\ && + 2\beta\dis
\frac{\partial}{\partial \beta}+2\gamma\dis
\frac{\partial}{\partial \gamma} \label{scaling1}
\end{eqnarray} This is a trivial finding.
\subsection{Eliminating $\beta$}
 We derive  $ \beta $ from the third equation of system (\ref{sisoriginario}),
 i.e.
$$\beta =-\dis\frac{ C \dot{r}}{mgx_{G}} $$
and obtain the following  system
 of four equations of first order, and one of second order:
 \begin{equation}
    \left\{
       \begin{array}{rcl}
          \ddot u_{1}&=&(mgx_{G}(u_{1}u_{4} -
          u_{3}u_{5})/{C}\\
          \dot u_{2}&=&(- Cu_{1}u_{3} + mg
          x_{G}u_{5})/{2C}\\
          \dot u_{3}&=&u_{1}u_{2}/{2}\\
          \dot u_{4}&=& (- Cu_{1}\dot
          u_{1}-mgx_{G}u_{2}u_{5})/{mg x_{G}}\\
          \dot u_{5}&=&(C \dot u_{1}u_{3} +{
          mg}x_{G}u_{2}u_{4})/{mgx_{G}}
      \end{array}
    \right.
    \label{sis1eo}
\end{equation} with \begin{equation} u_{1}={r},\;
u_{2}={q},\; u_{3}={p}, \;u_{4}=\alpha ,\; u_{5}=\gamma
\label{prima sost eo} \end{equation} If we apply Lie group
analysis to system (\ref{sis1eo}), then we obtain a determining
equation of parabolic type for $V$ in three independent variables.
Its two characteristic curves yield the following transformations
\begin{eqnarray} u_{4}=\dis\frac{s_{4}-
Cu_{1}^2}{2mg\,x_G} &{\rm viz} & \alpha =\dis\frac{s_{4}-
C{r}^2}{2mg\,x_G}\nonumber \\
u_{5}=\dis\frac{Cu_{3}u_{1} + s_{6}}{mg\,x_G} &{\rm viz} & \gamma
=\dis\frac{C{p r}+ s_{6}}{mg\,x_G}\label{cceo}
\end{eqnarray} with $s_{4}$
and $s_{6}$ new unknown functions of $ t $. Then, system   (\ref
{sis1}) transforms into:
 \begin{equation}
    \left\{
       \begin{array}{rcl}
          \ddot u_{1}&=&(- Cu_{1}^3 - 2Cu_{1}u_{3}^2 + u_{1}\tilde{u}_{4}
          - 2u_{3}\tilde{u}_{5})/{2C}\\
          \dot u_{2}&=&\tilde{u}_{5}/{2C}\\
          \dot u_{3}&=&{u_{1}u_{2}}/{2}\\
          \dot {\tilde{u}}_{4}&=&- 2u_{2}(Cu_{1}u_{3} + \tilde{u}_{5})\\
          \dot {\tilde{u}}_{5}&=&u_{2}( - 2Cu_{1}^2 + \tilde{u}_{4})/{2}
      \end{array}
    \right.
    \label{sis2eo}
\end{equation} with \begin{equation}  \tilde{u}_{4}=s_{4},\;\; \tilde{u}_{5}=s_{6}
\label{seconda sost eo} \end{equation} If we apply Lie group
analysis to system (\ref{sis2eo}), then we obtain two first order
partial differential equations for $ V $:
\begin{equation}
\dis\frac{\partial V}{\partial u_{3}} -4Cu_{3}\dis\frac{\partial
V}{\partial \tilde{u}_{4}}=0 \label{unoeo}
\end{equation} \begin{equation} \dis\frac{\partial
V}{\partial u_{2}} - 4Cu_{2}\dis\frac{\partial V}{\partial \tilde
u_{4}} =0 \label{dueeo}
\end{equation} with $ V \equiv V(u_2, u_3, \tilde u_4) $. From
(\ref{unoeo}) it is easy to obtain that $ V \equiv V(\eta, u_2) $
with
\begin{equation} \eta=2Cu_3^2+\tilde u_4 \label{eta} \end{equation}
Then, (\ref{dueeo}) becomes
\begin{equation} \dis\frac{\partial V}{\partial u_{2}}
-4Cu_{2}\dis\frac{\partial V}{\partial \eta}=0 \label{dueeo2}
\end{equation} Its characteristic curve is
\begin{eqnarray}
\xi_{1}=2Cu_2^2+\eta \label{xi1eo}
\end{eqnarray}
Finally, we have that $ V=\psi(\xi_1)$ with $\psi$ an arbitrary
function of $\xi$, and consequently operator
\begin{equation}
\Gamma_1=\psi(\xi_1){\partial}_t \label{opI1}\end{equation} is a
generator of a Lie point symmetry for system (\ref{sis2eo}).
Transforming (\ref{xi1eo}) into the original unknown functions by
using (\ref{eta}), (\ref{seconda sost eo}), (\ref{cceo}),
(\ref{prima sost eo}) yields
$$\xi_{1}=\dis\frac{C}{2}
\left( 2{p}^2+2{q}^2+{r}^2 \right)+mg\,x_G\alpha
$$
 which is exactly the first integral of conservation of energy (\ref{eo}).
 In addition, we have algorithmically derived  that
  (\ref{opI1}) is a generator of a Lie point symmetry for system
(\ref{sisoriginario}).
\subsection{Eliminating $\gamma$}
We derive  $ \gamma $ from the second equation of system
(\ref{sisoriginario}),
 i.e.
$$ \gamma =\dis\frac{C(2\dot {q}+{p
r})}{mgx_{G}}
$$
and obtain the following  system
 of four equations of first order, and one of second order:
\begin{equation}
    \left\{
       \begin{array}{rcl}
          \ddot u_{1}&=&u_{1}( - Cu_{3}^2 + 2mgu_{4}x_{G})/{4C}\\
          \dot u_{2}&=&u_{1}u_{3}/{2}\\
          \dot u_{3}&=&- mgu_{5}x_{G}/{C}\\
          \dot u_{4}&=&( - 2Cu_{1}\dot u_{1}
           - Cu_{1}u_{2}u_{3} + mgu_{3}u_{5}x_{G})/{mgx_{G}}\\
          \dot u_{5}&=&(2C\dot u_{1}u_{2}
          + Cu_{2}^2u_{3} - mgu_{3}u_{4}x_{G})/{mgx_{G}}
      \end{array}
    \right. \label{sis1kov}
\end{equation} with \begin{equation} u_{1}={q},\; u_{2}={p},\;
u_{3}={r},\; u_{4}=\alpha ,\; u_{5}=\beta \label{prima sost kov}
\end{equation} If we apply Lie group
analysis to system (\ref{sis1kov}), then we obtain a determining
equation of parabolic type for $V$ in three independent variables.
Its two characteristic curves yield the following transformations
\begin{eqnarray}
u_{4}=\dis\frac{s_{4}- Cu_{1}^2 }{mgx_{G}} &{\rm viz} & \alpha
=\dis\frac{s_{4}- C{q}^2}{mgx_{G}} \nonumber \\
u_{5}=\dis\frac{2Cu_{1}u_{2} + s_{5}}{mgx_{G}} &{\rm viz} & \beta
=\dis\frac{2C{p q}+ s_{5}}{mgx_{G}}\label {ccikov}
\end{eqnarray}  with $ s_{4} $ and $s_{5} $  new unknown functions of $ t $. Then, system
(\ref {sis1kov}) transforms into:
\begin{equation}
    \left\{
       \begin{array}{rcl}
          \ddot u_{1}&=&[u_{1}( - 2Cu_{1}^2 - Cu_{3}^2 + 2\tilde{u}_{4})]/4C\\
          \dot u_{2}&=&u_{1}u_{3}/2\\
          \dot u_{3}&=&(- 2Cu_{1}u_{2} - \tilde{u}_{5})/C\\
          \dot {\tilde{u}}_{4}&=&u_{3}(Cu_{1}u_{2} + \tilde{u}_{5})\\
          \dot {\tilde{u}}_{5}&=&u_{3}(Cu_{2}^2 - \tilde{u}_{4})\\
      \end{array}
    \right. \label{sis2kov}
\end{equation} with \begin{equation}
\tilde{u}_{4}=s_{4}, \;\;\;\tilde{u}_{5}=s_{5} \label{seconda sost
kov}
\end{equation}If we apply Lie group
analysis to system (\ref{sis2kov}), then we obtain two first order
partial differential equations for $ V $:
\begin{equation} u_3\dis\frac{\partial V}{\partial u_{2}}
-4u_{2}\dis\frac{\partial V}{\partial u_{3}}
\\ +4Cu_{2}u_3\dis\frac{\partial V}{\partial
\tilde{u}_{4}}=0  \label{unokov}
\end{equation}
\begin{equation}  8Cu_{2}\tilde{u}_5\dis\frac{\partial
V}{\partial u_{3}} -Cu_{3}\tilde{u}_{5} \dis\frac{\partial
V}{\partial \tilde{u}_{4}}+Cu_3 \left(\tilde{u}_{4}-Cu_{2}^2
\right)\dis\frac{\partial V}{\partial \tilde{u}_{5}} =0
\label{duekov}
\end{equation} with $ V \equiv V(u_{2},u_{3},\tilde{u}_{4},\tilde{u}_5)
$. From (\ref{unokov}) it is easy to obtain that $ V \equiv
V(\eta_1, \eta_2, \tilde{u}_5) $ with
\begin{equation} \eta_{1}=4u_2^2+u_3^2,\;\;\;\;  \eta_{2}=Cu_2^2-\tilde{u}_4\label{eta1}
  \end{equation}Then,
(\ref{duekov}) becomes \begin{equation}
2Cu_{3}\tilde{u}_5\dis\frac{\partial V}{\partial \eta_1}
+Cu_{3}\tilde{u}_{5} \dis\frac{\partial V}{\partial \eta_2} -Cu_3
\left(u_{4}-Cu_{2}^2 \right)\dis\frac{\partial V}{\partial
\tilde{u}_{5}} =0 \label{duekov2}
\end{equation}
 Its characteristic curves are
\begin{equation}
\xi_{1}=C\eta_{1}-2\eta_{2}, \;\;\;\;
\xi_{2}=\eta_{2}^2+\tilde{u}_5^2 \label{xi1kov}
\end{equation} Finally, we have that $ V=\Psi(\xi_1,\xi_2)$ with $\Psi$ an arbitrary
function of $\xi_1,\; \xi_2$, and consequently operator
\begin{equation}
\Gamma_1=\Psi(\xi_1,\xi_2){\partial}_t \label{opI2}\end{equation}
is a generator of a Lie point symmetry for system (\ref{sis2kov}).
Transforming (\ref{xi1kov}) into the original unknown functions by
using (\ref{eta1}), (\ref{seconda sost kov}), (\ref{ccikov}),
(\ref{prima sost kov}) yields
$$\xi_{1}=\dis\frac{C}{2}
\left( 2{p}^2+2{q}^2+{r}^2 \right)+mg\,x_G\alpha
$$
$$
\xi_{2}=\left({p}^2-{q}^2-\dis\frac{x_{G}\alpha mg}{C}\right)^2+
\left(2{p q}-\dis\frac{x_{G}\beta mg}{C}\right)^2 $$ which are
exactly the first integral of conservation of energy (\ref{eo}),
and
 that derived by Kowalevski (\ref{ikov}), respectively.
 In addition, we have algorithmically derived  that
  (\ref{opI2}) is a generator of a Lie point symmetry for system
(\ref{sisoriginario}).

\subsection{Eliminating $ p $}
We derive  $ p $ from the second equation of system
(\ref{sisoriginario}),
 i.e.
$$ {p}=\dis\frac{ mg\gamma
x_{G}- 2C\dot {{q}} }{C{r}}
$$ and obtain the following  system
 of four equations of first order, and one of second order:
\begin{equation}
    \left\{
       \begin{array}{rcl}
          \ddot u_{1}&=&u_{1}( - Cu_{3}^2 + 2mgu_{4}x_{G})/4C\\
          \dot u_{2}&=&(Cu_{1}u_{3}u_{4} + 2C\dot u_{1}u_{5} - mgu_{2}u_{5}x_{G})/Cu_{3}\\
          \dot u_{3}&=&- mgu_{5}x_{G}/C\\
          \dot u_{4}&=&- u_{1}u_{2} + u_{3}u_{5}\\
          \dot u_{5}&=&(- 2C\dot u_{1}u_{2} - Cu_{3}^2u_{4} +
          mgu_{2}^2x_{G})/Cu_{3}
      \end{array}
    \right. \label{sis1gen}
\end{equation} with \begin{equation} u_{1}={q},\; u_{2}=\gamma,\;
u_{3}={r},\; u_{4}=\alpha,\; u_{5}=\beta \label{prima sost gen}
\end{equation}  If we apply Lie group
analysis to system (\ref{sis1gen}), then we obtain a determining
equation of parabolic type for $V$ in three independent variables.
Its two characteristic curves yield the following transformations
\begin{eqnarray} u_{2}=\sqrt{s_{6}} \cos(2s_{5} - 2u_{3}) &{\rm viz} & \gamma
=\sqrt{s_{6}}\cos(2s_{5}-2{r}) \nonumber \\
u_{5}=\sqrt{s_{6}}\sin(2s_{5} - 2u_{3}) &{\rm viz} &  \beta
=\sqrt{s_{6}}\sin(2s_{5} - 2{r}) \label{ccgen}
\end{eqnarray} with $ s_{6} $  e $ s_{5} $  new unknown functions of $ t $. Then, system
(\ref{sis1gen}) transforms into:
\begin{equation}
    \left\{
       \begin{array}{ll}
          \ddot u_{1}=u_{1}( - Cu_{3}^2 + 2mgu_{4}x_{G})/4C\\
          \dot {\tilde{u}}_{2}=2u_{4}\sqrt{{\tilde{u}}_{2}}\cos(2u_{3} - 2{\tilde{u}}_{5})
          \Big[\tan(2u_{3} - 2{\tilde{u}}_{5})u_{3} + u_{1}\Big]\\
          \dot u_{3}=\sqrt{{\tilde{u}}_{2}}\sin(2u_{3} - 2{\tilde{u}}_{5})
          mgx_{G}/C\\
          \dot u_{4}=- \sqrt{{\tilde{u}}_{2}}\cos(2u_{3} - 2{\tilde{u}}_{5})
          \Big[\tan(2u_{3} - 2{\tilde{u}}_{5})u_{3} + u_{1}\Big]\\
          \dot {\tilde{u}}_{5}=\sqrt{{\tilde{u}}_{2}}\cos(2u_{3} -
          2{\tilde{u}}_{5})
          \Big[- (Cu_{3}^2u_{4} -mg{\tilde{u}}_{2}x_{G})- 2C\dot
           u_{1}{\tilde{u}}_{2}\\ \qquad \;\;\;+(Cu_{1}u_{4}+  2mg{\tilde{u}}_{2}x_{G})\tan(2u_{3} - 2{\tilde{u}}_{5})u_{3}
           \Big]/2C{\tilde{u}}_{2}u_{3}
      \end{array}
    \right. \label{sis2gen}
\end{equation} with \begin{equation}
\tilde{u}_{2}=s_{6}, \;\;\;\tilde{u}_{5}=s_{5} \label{seconda sost
gen}
\end{equation} If we apply Lie group
analysis to system (\ref{sis2gen}), then we obtain a determining
equation of parabolic type for $V$ in two independent variables.
Its characteristic curve yields the following transformation
\begin{eqnarray} \tilde{u}_{5}=\dis\frac{ss_{5} -  u _1}{u_{3}} &{\rm viz}
&
\begin{array}{ll}\gamma=\sqrt{s_6} \cos(2\dis\frac{ss_{5}-q-r^2}{r})
\\   \beta=\sqrt{s_6} \sin(2\dis\frac{ss_{5}-q-r^2}{r} )
\end{array}
\label{cc2gen}
\end{eqnarray} with $ss_{5}$ a new unknown function of $ t $. Then, system
 (\ref{sis2gen}) transforms into
\begin{equation}
    \left\{
       \begin{array}{ll}
          \ddot u_{1}=u_{1}( - Cu_{3}^2 +
          2mgu_{4}x_{G})/4C\\
          \dot {\tilde{u}}_{2}=2u_{4}\sqrt{\tilde u_{2}}\cos\Big((2u_{1} + 2u_{3}^2
          - 2{\hat{u}}_{5})/u_{3}\Big)
          \Big [u_1\\ \qquad \;\;\;\;\;\;\;\;\;\;\;\;\;\;\;\;\;\;\;\;\;\;+\tan\Big((2u_{1} + 2u_{3}^2
          - 2{\hat{u}}_{5})/u_{3}\Big )u_{3} \Big ]\\
          \dot u_{3}=\sqrt{\tilde u_{2}}\sin\Big((2u_{1} + 2u_{3}^2 - 2{\hat{u}}_{5}
          )/u_{3}\Big)mgx_{G}/C\\
          \dot u_{4}=- \sqrt{\tilde u_{2}}\cos\Big((2u_{1} + 2u_{3}^2 - 2{\hat{u}}_{5}
          )/u_{3}\Big)\Big[u1\\ \qquad \;\;\;\;\;\;\;\;\;\;\;\;\;\;\;\;\;\;\;\;\;\;
          +\tan\Big((2u_{1} + 2u_{3}^2
          - 2{\hat{u}}_{5})/u_{3}\Big )u_{3} \Big ]\\
          \dot {\hat{u}}_{5}=\sqrt{\tilde u_{2}}
          \cos\Big(\left(2u_{3}^2 - 2{\hat{u}}_{5} + 2u_{1}\right)/u_{3}\Big)
           \bigg[\Big[(2u_{3}^2 +
          2{\hat{u}}_{5}\\ \qquad \;\;\; - 2u_{1})mg\tilde u_{2}x_{G} +
          Cu_{1}u_{3}^2u_{4}\Big]
          \tan\Big((2u_{3}^2 - 2{\hat{u}}_{5} +
          2u_{1})/u_{3}\Big)\\ \qquad \;\;\;\;+
          (mg\tilde u_{2}x_{G}-cu_{3}^2u_{4})u_{3}\bigg]/2c\tilde u_{2}u_{3}
      \end{array}
    \right. \label{sis3gen}
\end{equation} with \begin{equation}
\hat{u}_{5}=ss_{5} \label{terza sost gen}
\end{equation}If we apply Lie group
analysis to system (\ref{sis3gen}), then we obtain one first order
partial differential equation for $ V $:
 \begin{equation}  \dis\frac{\partial
V}{\partial u_{4}} -2u_{4} \dis\frac{\partial V}{\partial \tilde
u_{2}}=0 \label{unogen}
\end{equation} with $ V \equiv V(\tilde u_{2},u_{4})
$. Its characteristic curve is
\begin{eqnarray}
\xi_{1}=\tilde u_{2}+u_{4}^2 \label{xi1gen}
\end{eqnarray} Finally, we have that $ V=\psi(\xi_1)$ with $\psi$ an arbitrary
function of $\xi_1$, and consequently operator\begin{equation}
\Gamma_1=\psi(\xi_1){\partial}_t \label{opI3}\end{equation} is a
generator of a Lie point symmetry for system (\ref{sis3gen}).
Transforming (\ref{xi1gen}) into the original unknown functions by
using (\ref{terza sost gen}), (\ref{cc2gen}), (\ref{seconda sost
gen}), (\ref{ccgen}), (\ref{prima sost gen}) yields
$$\xi_{1}=\alpha^2+\beta^2+\gamma^2 $$ which is exactly the first integral
of the length of the unit vertical vector (\ref{igenerale}).
 In addition, we have algorithmically derived  that
  (\ref{opI3}) is a generator of a Lie point symmetry for system
(\ref{sisoriginario})

\section{Kepler problem}
In \cite{harmony}, Nucci's method  \cite{kepler} was used to find
symmetries additional to those reported by Krause \cite{Krause} in
his study of the complete symmetry group of the Kepler problem.  A
consequence of the application of Nucci's method was the
demonstration of the group theoretical relationship between the
simple harmonic oscillator and the Kepler
 problem. In \cite{harmony}, polar coordinates were used, and Nucci's method was not
 applied to the three-dimensional case with the purpose of finding first integrals.
 We do it here by considering cartesian coordinates.\\ The
 equations of motion of the Kepler problem are given by the following
  well-known three equations of second order
\begin{equation}
  \left\{
    \begin{array}{rcl}
       \ddot x_{1}&=& -\mu x_{1}/\Big((x_{1}^2+x_{2}^2+x_{3}^2)^{3/2}\Big)\\
       \ddot x_{2}&=& -\mu x_{2}/\Big((x_{1}^2+x_{2}^2+x_{3}^2)^{3/2}\Big)\\
       \ddot x_{3}&=& -\mu x_{3}/\Big((x_{1}^2+x_{2}^2+x_{3}^2)^{3/2}\Big)\\
    \end{array}
 \right.
\label {eqkep} \end{equation} The first integrals for the Kepler
problem are: conservation of energy $E$, conservation of angular
momentum ${\mathbf K}$, the Laplace-Runge-Lenz vector ${\mathbf
L}$. None of the unknowns $x_1,x_2,x_3,\dot x_1,\dot x_2, \dot
x_3$ are missing in the expression of $E$ and the components of
${\mathbf L}$. This is not true for the three components of
${\mathbf K}$, i.e.
\begin{eqnarray}
K_1 &=& x_3 \dot x_2- \dot x_3 x_2 \label{momqdimoto1} \\
K_2 &=& x_3 \dot x_1- \dot x_3 x_1 \label{momqdimoto2}\\
K_3 &=& x_1 \dot x_2- \dot x_1 x_2 \label{momqdimoto3}
\end{eqnarray}
Therefore,
 we can only obtain the three components of ${\mathbf K}$ using
our method. However, neither $E$ nor ${\mathbf L}$ are needed to
reduce system (\ref{eqkep}) to a linear oscillator, as we show in
the following. Let us transform system (\ref {eqkep}) into a
system of six equations of first order
\begin{equation}
  \left\{
    \begin{array}{rcl}
       \dot w_{1}&=&w_{4}\\
       \dot w_{2}&=&w_{5}\\
       \dot w_{3}&=&w_{6}\\
       \dot
       w_{4}&=&-\mu w_{1}/\Big((w_{1}^2+w_{2}^2+w_{3}^2)^{3/2}\Big)\\
       \dot
       w_{5}&=&-\mu w_{2}/\Big((w_{1}^2+w_{2}^2+w_{3}^2)^{3/2}\Big)\\
       \dot
       w_{6}&=&-\mu w_{3}/\Big((w_{1}^2+w_{2}^2+w_{3}^2)^{3/2}\Big)
    \end{array}
 \right.
\label{eqkep1} \end{equation}  with \begin{equation}
w_{1}=x_{1},\; w_{2}=x_{2},\; w_{3}=x_{3},\; w_{4}=\dot x_{1},\;
w_{5}=\dot x_{2},\;
 w_{6}=\dot x_{3} \label{prima sost} \end{equation}
 Consequently, the components of the angular
momentum become
\begin{eqnarray}
K_1=w_3w_5-w_6 w_2 \label{w5} \\
K_2=w_3w_4 - w_1w_6 \label{w4} \\
K_3=w_1w_5-w_4w_2 \label{w2}
\end{eqnarray}
 We choose one of the dependent variables to be the new independent
variable $y$ in order to reduce  the order of system
(\ref{eqkep1}) by one \cite{kepler}. We take $w_3=y$.  Then,
system (\ref{eqkep1}) becomes the following non-autonomous system
of five first order ordinary differential equations
\begin{equation}
    \left\{
        \begin{array}{rcl}
            w_{1}'&=&w_{4}/w_{6}\\
            w_{2}'&=&w_{5}/w_{6}\\
            w_{4}'&=& - \mu w_{1}/\Big(w_{6}(w_{1}^2 + w_{2}^2 +
            y^2)^{3/2}\Big)\\
            w_{5}'&=&- \mu w_{2}/\Big(w_{6}(w_{1}^2 + w_{2}^2 +
            y^2)^{3/2}\Big)\\
            w_{6}'&=& -\mu y/\Big(w_{6}(w_{1}^2 + w_{2}^2 +
            y^2)^{3/2}\Big)
        \end{array}
     \right.
\label{sist originario} \end{equation} with $'$ denoting
differentiation with respect to $y$. Let us observe that
\begin{itemize}
\item
 $w_4$ does not appear in $K_1$
\item $w_5$ does not appear in $K_2$
\item $w_6$ does not appear in $K_3$
\end{itemize}
We should remark that other variables are missing too. For
example, $w_1$ is also missing in $K_1$. However, our method will
yield the result whatever the choice of a missing variable. In the
following, we eliminate $w_4,w_5,w_6$ from system (\ref{sist
originario}) one at a time.
\subsection{Eliminating $w_4$}
We derive $w_4$
from the first equation of system(\ref{sist originario}), i.e.
$$ w_{4}=w_{1}'w_{6} $$ and obtain the following  non-autonomous system of
three equations of first order, and one of second order:
\begin{equation}
   \left\{
     \begin{array}{rcl}
        u_{4}''&=& \mu ( u_{4}'y- u_{4} )/\Big(u_{3}^2(u_{2}^2 + u_{4}^2 +
        y^2)^{3/2}\Big)\\
        u_{3}'&=&- \mu y/\Big(u_{3}(u_{2}^2 + u_{4}^2 +
        y^2)^{3/2}\Big)\\
        u_{2}'&=&u_{1}/u_{3}\\
        u_{1}'&=&- \mu u_{2}/\Big(u_{3}(u_{2}^2 + u_{4}^2 + y^2)^{3/2}\Big)
      \end{array}
   \right. \label{ultimo sis1}
\end{equation} with \begin{equation} u_{4}=w_{1},\;
u_{2}=w_{2},\; u_{3}=w_{6},\;  u_{1}=w_{5} \label{tutte le u 1}
\end{equation}
If we apply Lie group analysis to system (\ref{ultimo sis1}), then
after several reductions we obtain one first order partial
differential equations for $ G_3 $:\begin{equation}
u_1\dis\frac{\partial G_3}{\partial u_{2}}
+u_{3}\dis\frac{\partial G_3}{\partial y}=0 \label{cc1comp}
\end{equation}
with $ G_3 \equiv G_3(u_{1},u_{2},u_{3},y) $. Its solution is $
G_3=\psi(\xi_1)$ with $\psi$ an arbitrary function of
\begin{equation} \xi_1=u_3u_2-yu_1
\label{prima sost della comp} \end{equation} Transforming
(\ref{prima sost della comp}) into the original unknown functions
by using (\ref{prima sost}), (\ref{tutte le u 1}) yields
$$ \xi_1=\dot x_{3} x_{2} - \dot x_{2} x_{3} $$
 which is exactly the first component
 of the angular momentum (\ref{momqdimoto1}).

\subsection{Eliminating $w_5$} We derive $w_5$ from the second
equation of system (\ref{sist originario}), i.e.
$$w_{5}=w_{2}'w_{6} $$ and obtain the following
 non-autonomous system of three equations of first order, and one of second
order:
\begin{equation}
   \left\{
     \begin{array}{rcl}
        u_{4}''&=& \mu ( - u_4 + u_4'y)/\Big(u_2(u_1^2 + u_4^2 +
        y^2)^{3/2}\Big)\\
        u_{3}'&=&- \mu u_1/\Big(u_2(u_1^2 + u_4^2 +
        y^2)^{3/2}\Big)\\
        u_{2}'&=&- \mu y/\Big((u_1^2 + u_4^2 +
        y^2)^{3/2}\Big)\\
        u_{1}'&=&u_3/u_2
      \end{array}
   \right. \label{ultimo sis2}
\end{equation} with \begin{equation} u_{4}=w_{2},\;
u_{2}=w_{6},\; u_{3}=w_{4},\;  u_{1}=w_{1} \label{tutte le u 2}
\end{equation}
If we apply Lie group analysis to system (\ref{ultimo sis2}), then
after several reductions we obtain one first order partial
differential equation for $ G_2 $:\
\begin{equation} u_3\dis\frac{\partial
G_2}{\partial u_{1}} +u_{2}\dis\frac{\partial G_2}{\partial y}=0
\label{cc2comp}
\end{equation}
with $ G_2 \equiv G_2(u_{1},u_{2},u_{3},y) $. Its solution is $
G_2=\phi(\xi_2)$ with $\phi$ an arbitrary function of
\begin{equation} \xi_2=yu_3-u_{1}u_{2}
\label{prima sost della comp2} \end{equation}  Transforming
(\ref{prima sost della comp2}) into the original unknown functions
by using (\ref{prima sost}), (\ref{tutte le u 2}) yields
$$ \xi_2=x_{3} \dot x_{1} - \dot x_{3} x_{1} $$ which is exactly the
second component
 of the angular momentum  (\ref{momqdimoto2}).

\subsection{Eliminating $w_6$} We derive $w_6$ from the first
equation of system(\ref{sist originario}), i.e.
$$w_{6}=\dis\frac{w_{4}}{w_{1}'} $$ and obtain the following
non-autonomous system of three equations of first order, and one
of second order:
\begin{equation}
   \left\{
     \begin{array}{rcl}
        u_{4}''&=& \mu  u_4'^2( - u_4 + u_4'y)/\Big(u_1^2(u_2^2 + u_4^2 +
        y^2)^{3/2}\Big)\\
        u_{3}'&=&- \mu u_2u_4'/\Big(u_1(u_2^2 + u_4^2 +
        y^2)^{3/2}\Big)\\
        u_{2}'&=&u_3u_4' /u_1\\
        u_{1}'&=&- \mu u_4u_4' /\Big(u_1(u_2^2 + u_4^2 + y^2)^{3/2}\Big)
     \end{array}
   \right. \label{ultimo sis3}
\end{equation} with \begin{equation} u_{4}=w_{1},\;
u_{2}=w_{2},\; u_{3}=w_{5},\;  u_{1}=w_{4} \label{tutte le u 3}
\end{equation}
If we apply Lie group analysis to system (\ref{ultimo sis3}), then
after several reductions we obtain one first order partial
differential equation for $ G_3 $:
\begin{equation} u_3\dis\frac{\partial G_3}{\partial u_{2}}
+u_{1}\dis\frac{\partial G_3}{\partial u_1}=0 \label{cc3comp}
\end{equation}
with $ G_3 \equiv G_3(u_{1},u_{2},u_{3},u_4)$ with $ G_3 \equiv
G_3(u_{1},u_{2},u_{3},y) $. Its solution is $ G_3=\varphi(\xi_3)$
with $\varphi$ an arbitrary function of
\begin{equation} \xi_3=u_1u_2-u_{4}u_{3}
\label{prima sost della comp3} \end{equation} Transforming
(\ref{prima sost della comp3}) into the original unknown functions
by using (\ref{prima sost}), (\ref{tutte le u 3}) yields
$$ \xi_3=\dot x_{2} x_{1} - \dot x_{1}
x_{2}$$ which is exactly the third component
 of the angular momentum (\ref{momqdimoto2}).

Now let us derive $w_5$, $w_4$ and $w_2$ from (\ref{w5}),
(\ref{w4}) and (\ref{w2}), i.e.
\begin{eqnarray}
w_5&=&\dis\frac{-\xi_3w_6 y + \xi_2\xi_1 +
\xi_1w_1w_6}{\xi_2 y} \label{w5p} \\
w_4&=&\dis\frac{\xi_2 + w_1w_6}{y} \label{w4p} \\
w_2&=&\dis\frac{- \xi_3 y + \xi_1w_1}{\xi_2} \label{w2p}
\end{eqnarray}
with $\xi_1$, $\xi_2$, $\xi_3$ new unknown functions of $y$.
Substituting  (\ref{w5p}), (\ref{w4p}), (\ref{w2p}) into
(\ref{sist originario}), and deriving  $ w_6 $ from the first
equation yields the following system of three equations of first
order, and one of second order:
\begin{equation}
 \left\{
     \begin{array}{rcl}
      u_4''&=&\Big(\mu u_2( - u_4^3 + 3u_4^2u_4'y - 3u_4u_4'^2y^2 +
      u_4'^3y^3)\Big)/  \Big((u_1^2y^2 \\ && \qquad - 2u_3u_1u_4y + u_2^2u_4^2 + u_2^2y^2 +
      u_3^2u_4^2)^{3/2}\Big) \\
      u_3'&=&0\\
      u_2'&=&0\\
      u_1'&=&0 \end{array}
   \right.\label{u4}
\end{equation}
with $$ u_4=w_1,\; u_3=\xi_3,\; u_2=\xi_2,\; u_1=\xi_1$$ It is
easy to show that system (\ref{u4}) admits an eleven-dimensional
Lie symmetry algebra. In fact, the first equation of (\ref{u4})
itself admits a Lie symmetry algebra of dimension eight, which
means that it is is linearizable through a point transformation
\cite{Lie 5}. Thus, we have reduced the equations of motion of the
Kepler problem to the harmonic oscillator \cite{harmony},
\cite{classic} by using Lie group analysis.

\section{Final comments}

We have found that Lie group analysis yields the first integrals
admitted by any system of ordinary differential equations if the
method developed by Nucci \cite{kepler} is applied, the only
limitation being the absence of at least  one of the unknowns in
each first integral. \\Is it possible to obtain all of the first
integrals by means of Lie group analysis?  Also, what is the link
between Painlev\'e method and Lie group analysis \cite{kruskal}?
In addition, can Lax pairs be found by Lie group analysis? So far
these
are open questions that we hope to address in future work.\\
Let us conclude by underlining  that the application of Nucci's
method to the Kowalevski top  have led us to understand how first
integrals can be found by using Lie group analysis.
 In 1984, Cooke \cite{Cooke} wrote ``Kowalevskaya's work
is an ingenious application of mathematics to a system of
equations of great mathematical interest $\ldots$ but since the
case to which it applies is rather special, the details of her
arguments are no longer worth troubling about". About the same
time, a revival of interest into integrable problems of mechanics
has led to numerous papers on the Kowalevski top. Just to cite a
few, see \cite{Peremolov}, \cite{Hermann}, \cite{HaiH},
\cite{AdlM}, \cite{BobRS}, \cite{KomK} and the entire no. 11 issue
of the Journal of Physics A
 {\bf 34}, (2001).

\end{document}